# Initial Indications of Safety of Driverless Automated Driving Systems

Jiayu Joyce Chen
Department of Civil and Environmental Engineering
University of California, Berkeley
Email: chenjiayu@berkeley.edu

Steven E. Shladover
California PATH Program,
University of California, Berkeley,
Email: steve@path.berkeley.edu

Word Count: [6160] words + [4] tables (250 words per table) = [7160] words

## Abstract

As driverless automated driving systems (ADS) start to operate on public roads, there is an urgent need to understand how safely these systems are managing real-world traffic conditions. With data from the California Public Utilities Commission (CPUC) becoming available for Transportation Network Companies (TNCs) operating in California with and without human drivers, there is an initial basis for comparing ADS and human driving safety.

This paper analyzes the crash rates and characteristics for three types of driving: Uber ridesharing trips from the CPUC TNC Annual Report in 2020, supervised autonomous vehicles (AV) driving from the California Department of Motor Vehicles (DMV) between December 2020 and November 2022, driverless ADS pilot (testing) and deployment (revenue service) program from Waymo and Cruise between March 2022 and August 2023. All of the driving was done within the city of San Francisco, excluding freeways. The same geographical confinement allows for controlling the exposure to vulnerable road users, population density, speed limit, and other external factors such as weather and road conditions. The study finds that supervised AV has almost equivalent crashes per million miles (CPMM) as Uber human driving, the driverless Waymo AV has a lower CPMM, and the driverless Cruise AV has a higher CPMM than Uber human driving. The data samples are not yet large enough to support conclusions about whether the current automated systems are more or less safe than human-operated vehicles in the complex San Francisco urban environment.

## Introduction

The development of Automated Driving Systems (ADS) is changing the transportation industry. It offers the potential to free humans from driving tasks so that they can focus on other tasks, as well as to provide mobility to people who would otherwise be unable to travel. However, concerns remain about the safety of the deployment of vehicles driven by ADS (more commonly described as automated vehicles or AVs) on public roads, particularly regarding their interactions with other road users.

The International Organization for Standardization (ISO) defines automotive-specific functional safety to be an "absence of unreasonable risk" (1). There are currently no articles in Federal Motor Vehicle Safety Standards (FMVSS) to govern AV safety in the U.S. because this is determined by the driving behavior of the ADS software, which falls into a gray zone between traditional federal and state regulatory roles. Consequently, it has been necessary for states to step into the breach to provide some degree of protection for public safety. In California, three categories of reporting requirements are relevant to ADS operations, from the National Highway Traffic Safety Administration (NHTSA), California Public Utilities Commission (CPUC), and California Department of Motor Vehicles (CA DMV), as shown in Figure 1 below.

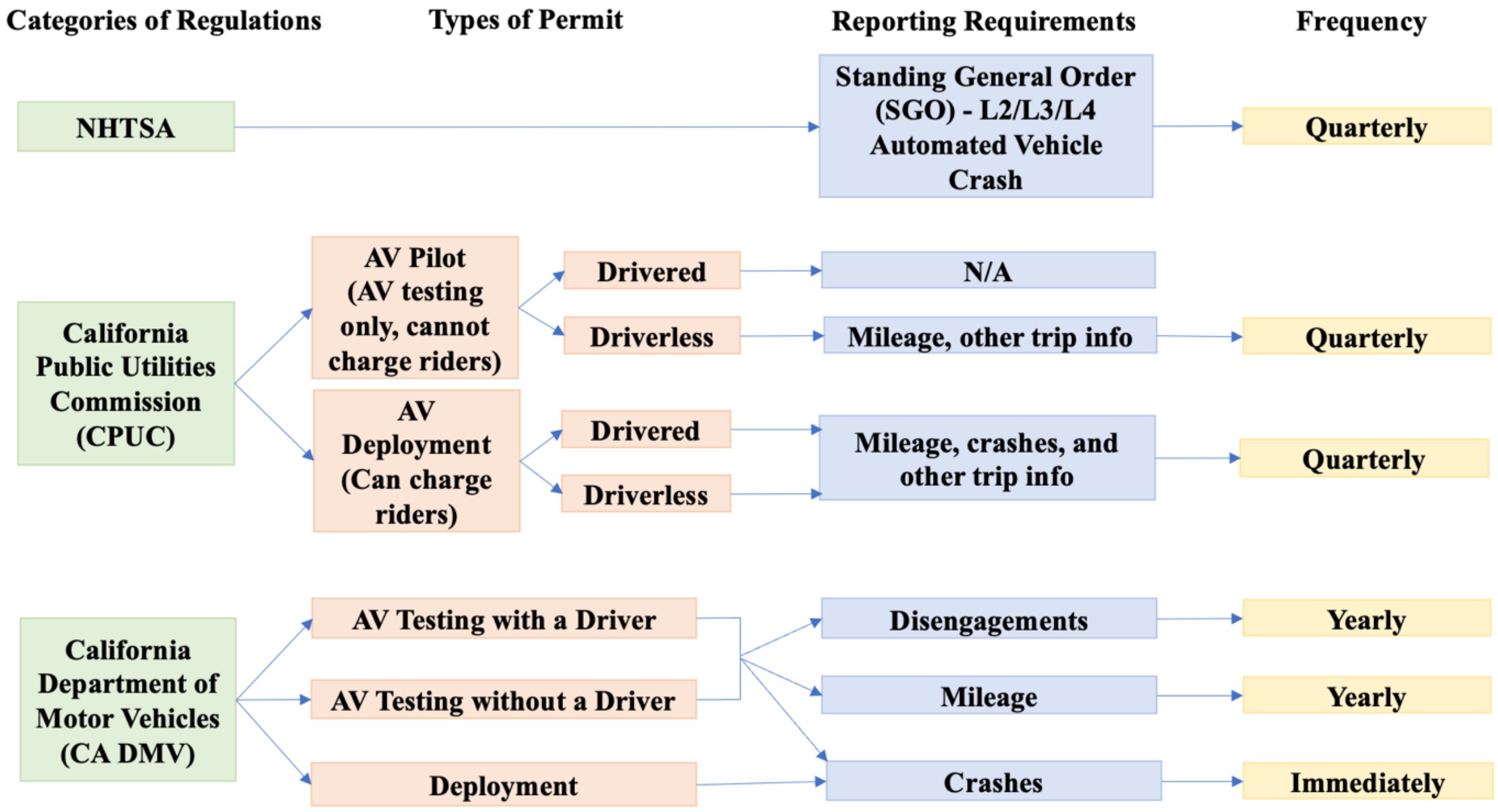

Figure 1- Categories of ADS vehicle driving regulation, and their respective types of permits, reporting requirements, and frequency.

NHTSA is not a permitting agency, however, it has issued a Standing General Order (SGO) requiring manufacturers and AV operators to "report to the agency certain crashes involving vehicles equipped with automated driving systems or SAE Level 2 advanced driver assistance systems" (2). However, it does not require reporting of the mileage driven for the AV operators. Hence, with only NHTSA's data, the frequency of occurrence of these crashes per mile is unknown.

In California, the CA DMV and CPUC both have jurisdiction over the AV. The DMV issues testing and deployment permits under a three-stage sequence of regulations to govern the safety of AV operations on public roads at three stages of development (3):

(a) Testing under the constant supervision of an in-vehicle safety driver;
(b) Testing without an in-vehicle safety driver, but with remote supervision;
(c) Commercial operation.

During the testing stages, CA DMV requires the AV companies to file reports on the mileage driven, disengagements, and crashes, as surrogate assessments for AV safety. The disengagement reporting requirement was intended to acquire information about the frequency of occurrence of various types of problems, both internal and external to the system, that makes it impossible for the ADS to complete the dynamic driving task. In these cases, the safety driver would identify a problematic situation and take over the dynamic driving task. Unfortunately, the disengagement reporting requirement was not defined sufficiently precisely to lead to uniform and comprehensive reporting of all the safety-relevant disengagements that occur. In the deployment stage, the AV companies are only required to report crashes, and so they ceased to report disengagements and mileage. Since crashes are rare events, this data sample has been sparse until very recently, when the mileage of driverless AV testing and operations has been growing.

The deployment permit from CA DMV is a prerequisite to the CPUC's permit, which is a requirement "for companies that provide transportation services to the public using AVs" (3). There are two categories of operation, Pilot, when the AV can offer rides to passengers but cannot charge the riders, and Deployment, when the AV can offer rides to passengers and charge them fares. Each of these categories has two types of permits: drivered, when the AV has a driver, and driverless, when there is no driver. The most recent approval of permits on August 10$^{th}$, 2023, for AVs to operate driverless deployment services, along with the numerous safety-related incidents, have sparked a local public outcry.

So, now more than ever, it is critical for researchers to study the crash reports to assess the progress and safety of these AVs. This paper focuses on the initial limited data available for comparing crashes of AV operations to human driving crashes in the City and County of San Francisco, which has been the primary host site for driverless activity in California.

## Literature Review

The current literature has focused on analyzing the causes and characteristics of AV crashes from California DMV AV reports (4), and assessing crashes and disengagements together as safety indicators.

Wang et al. (5) examined AV crashes on public roads from 2014-2018. Over 3.7 million miles were driven, and disengagement frequency varied significantly for different manufacturers. The paper found that 94% of the crashes were caused by other parties and proposed that AVs should proactively alert the human operator to avoid safety risks caused by the other parties.

Das et al. (6) used a Bayesian latent class model to identify six classes that correlated based on different variables and collision traits for AVs. It found that injury severity level was correlated with turning, multi-vehicle collisions, sideswipe and rear-end collisions, and dark lighting conditions with streetlights. The authors also urged advanced and robust collision narrative reporting to better understand and evaluate the collision likelihoods of AVs. Similarly, Xu et al. (7) used bootstrap-based binary logistic regression models to find that driving mode, collision location, roadside parking, rear-end collision, and one-way roads are significant in determining the severity level. The authors recommended that the DMV standardize data, such as reporting for severity level of crashes, for future use.

Song et al. (8) examined the sub-sequences leading up to a crash and found that "collision after AV stopped" was a common phenomenon. It also concluded that 68% of disengagements occurred in situations that would have led to an imminent collision absent the disengagement. The paper recommended using behavior and characteristics leading up to a crash for AV scenario testing.

There are also findings available from the companies that develop ADS. From Waymo's study by Victor et al (9), the company did not have any reported injuries within their first one million miles of driverless operation. Waymo had twenty contact events, two of which were severe enough to meet the criteria for the National Highway Traffic Safety Administration's Crash Investigation Sampling System, and 18 were minor crashes. In addition, Waymo reported that 55% of their crashes occurred when their AV was stationary, which means the fault was with the other vehicle driver. Waymo has also published a study (10) that used simulations to reconstruct 72 fatal crashes between 2008-2017 in the Chandler,

Arizona area where they were operating extensively, and substituted the human drivers with the Waymo AV software in the simulations. Their simulation results showed that the Waymo Driver avoided or mitigated 100% of the crashes except for the cases when the AV was struck from behind. The paper also highlighted that they were able to avoid all 20 instances that involved a pedestrian or cyclist.

Waymo has also collaborated with Swiss Re to compare their AV insurance claim frequency with private passenger vehicle claim rates in San Francisco CA and the Phoenix AZ metro area (11). Waymo used their own AV's third-party liability claims, which arise following damage to property or injury after a crash, with mileage that the AV has operated (separately for cases with and without safety drivers). Swiss Re provided privately owned vehicle insurance claims in the same zip codes as Waymo services and estimated the crash rates using estimated annual vehicle miles traveled by the insured drivers in these areas. The study found that driverless operations had 0 injuries (100% reduction as compared to human driving) and had 0.78 property damage claims per million miles (CPMM) as compared to 3.26 CPMM from human drivers. However, the study used estimated annual miles traveled by insured human drivers to create a baseline, which may be inaccurate. It also combined the San Francisco and Phoenix data, leaving it impossible to determine how different the results are for these two very different urban areas.

Cruise similarly has published a blog post that compared their AV's crash rate during its first million miles driven to that of over 5.6 million miles of human ride-hail driving in San Francisco (12). They claimed to have "65% fewer collisions overall, 94% fewer collisions as the primary contributor, and 74% fewer collisions with meaningful risks of injury". The numbers presented in the blog post were explained in a related paper from the University of Michigan (13). The paper describes two studies, both conducted with leased ride-hail vehicles in San Francisco. One study was conducted by the University of Michigan Transportation Research Institute (UMTRI) with 1149 vehicles operating for over 5 million miles, and the other was conducted by Virginia Tech Transportation Institute (VTTI) with 89 vehicles operating for around half a million miles. Some of the vehicles were studied by both VTTI and UMTRI, so the authors used a Bayesian Fusion statistical model and some non-conservative assumptions about discrepancies between the crash estimates from the two analysis methods to combine the two datasets to calculate a human-driven crash rate of 1 crash per 15,414 miles. The study also estimated the crash rate with meaningful risk of injury (which they assumed to be for any impact speed of 5 mph or more) is 1 per 85,027 miles. However, the study had some issues: first, it may have overestimated the crashes by potentially double counting crashes that did not exactly match up in the two datasets but may be the same crash. This would make the crash rate of ride-hail driving look higher than it is. Second, the process of identifying crashes within the San Francisco geofenced area is unclear, the authors identified a crash rate of "high confidence" in a small area of SF. However, they later chose a different and higher crash rate to compute crashes happening within all of SF. This would increase the count of crashes and again make the rate of ride-hail crashes appear higher.

The sensation that safety should be critically measured against human driving is shared by academia. Many scholars and experts recommend that the way to evaluate AV safety would be to look at exposure to risk from AV and compare it to that of human drivers. Blumenthal et al (14) explore different approaches to safety: safety as codified concepts for AV, safety as a measurement of injury/property damage, etc., safety as a process and regulation, and finally safety as a threshold (which uses human driving performance as a safety benchmark for AV). A similar idea is explored by Koopman (15), that the liability of AV in case of a crash would be determined by whether an attentive and unimpaired "reasonable human driver" would be able to resolve the situation. Their team recommends that "if the automated vehicle imitates the risk mitigation behaviors of the hypothetical reasonable human driver, no liability attaches for AV performance". It's unclear from the paper how the evaluation can be done in practice.

The literature review shows that there are some gaps in establishing an apples-to-apples comparison between AV crashes and human-driven car crashes. This study contributes to narrowing this gap by examining publicly available data reported to the State of California on miles driven by TNCs and AVs in San Francisco and comparing the characteristics of their respective reported crash cases. The dataset allows us to establish direct comparisons as the AVs and TNC vehicles are operating in similar urban conditions and have submitted data on these events to the same regulatory agency.

## Data and Methodology

### Data Selection

The analysis approach is constrained by the limited availability of relevant data. The data challenges exist on both the AV side and on the side of the human driving baseline for comparison. The reporting requirements on the AV companies are limited in both data elements and timeliness of updates. For the human driving baseline, the challenges are associated with identifying the existing data that can shed light on the safety records of human drivers operating under the same conditions as the AVs. This is a particular challenge for making the comparison with the driverless AV testing and commercial operations that are occurring in automated ride-hailing on many of the congested urban streets (but not freeways or the highest speed arterials) within the San Francisco city limits.

The following datasets were used for making the comparison:

1. AV Mileage Report data from DMV (16): The DMV publishes all reported miles driven in both supervised and driverless AVs annually for each company, which are tabulated separately based on which permit is involved. Supervised miles record automated miles driven by the AV, but the AV had safety drivers in the car who could take over control in difficult situations. Driverless means that the AV does not have any driver inside the vehicle, although it is still being monitored by remotely-located human assistants.
2. AV Collision Report data from DMV (17): The DMV publishes all reported AV-involved crashes that resulted in property damage, bodily injury, or death in California.
3. California Public Utilities Commission (CPUC) Transportation Network Company (TNC) Annual Report Data (18): The TNCs (Uber and Lyft, for example) are required to report annually on data such as Accessibility, Accidents & Incidents, Assaults & Harassments, Number of Hours, Driver Training, Requests Accepted, etc. This dataset was selected because we can filter for miles driven and crash data by zip code. This allows us to analyze human driving in the same areas where the AVs are operating, in particular the City and County of San Francisco (which have the same geographical boundary). The road network in San Francisco has mostly local roads with speed limits of 25 or lower mph. The CPUC data also requires crash reporting, which provides the number of crashes, given the miles driven. The crash reporting contains characteristics of the crash, which makes a comparison between the two datasets possible.
4. CPUC AV Program Quarterly Reporting, AV Pilot Program (19): The AV Pilot Program allows AV companies to offer unpaid driverless rides to the public. In return, they are required to report quarterly on the miles driven pertaining to this permit.
5. CPUC AV Program Quarterly Reporting, AV Deployment Program (19): The AV Deployment Program allows AV companies to offer paid driverless rides to the public. The companies are required to report quarterly on the miles driven and number of passenger trips, as well as crash occurrences, similar to the CPUC TNC reported data.
6. NHTSA Standing General Order on Crash Reporting (2): The AV companies are required to report all crashes, regardless of severity level, to NHTSA. The attributes of the crash data show whether the vehicle had a driver, what city it had crashed in, etc. This allows us to filter the data and identify those of interest.

### Data Processing and Categorization

#### Supervised Autonomous Mileage Data from DMV, Dec 2019 - Nov 2022

The supervised AV mileage came with the annual Disengagement Reporting from DMV (16). It tabulated licensed AVs from each company, and the automated miles they have driven monthly. We decided to use the data from December 2019 to November 2022 to match the CPUC TNC Annual Reports date available range. It also allows capturing AV trends since the start of the COVID-19 pandemic.

**Supervised Autonomous Vehicle Collision Report data from DMV, Dec 2019 - Nov 2022**
The AV crash data is documented from the crash reports, provided in the format shown in the following Figure 2 (16). Likewise, we only studied AV crashes between December 2019 and November 2022.

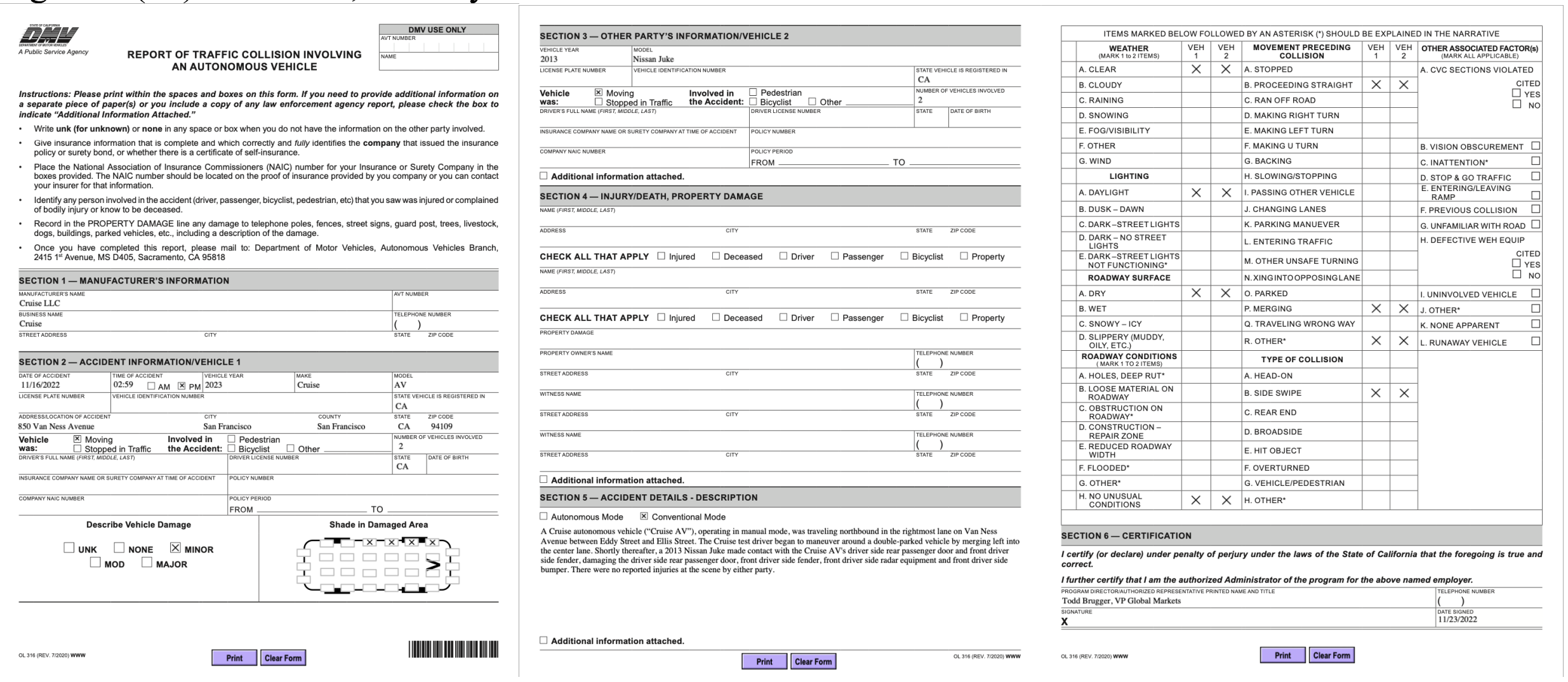

Figure 2 - Sample AV Crash Report Provided by the DMV

Each report is first converted into an entry in the table that we create to store all supervised AV crashes and their attributes. We used computer vision techniques to scan the texts and the checkboxes in the report, and the data were verified manually. Through the City field on the reports, we were able to identify the crashes that occurred in San Francisco. The steps are shown in Figure 3 below.

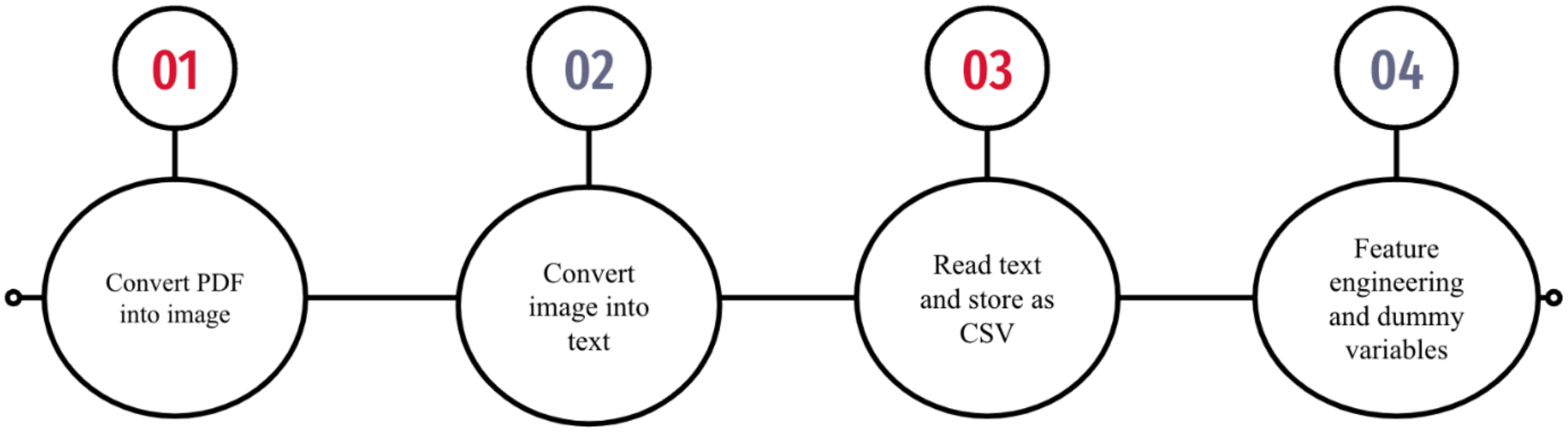

Figure 3 – Diagram of the PDF Data Processing Steps

Figure 3 shows the steps of data processing. We first convert the PDF reports into image files, so that the Python library can read the text in the image. Then we use the Optical Character Recognition and PDFMiner libraries to read the text and markings of the report. After the texts and checkmarks are extracted from the report, they are written into a CSV file with the pre-set columns representing the attributes. The data are validated with manually labeled data. Finally, we alter some of the formats of the data to allow better analysis, such as creating dummy variables for weather, lighting conditions, and road types.

Through data analysis, we further identified two categories of data that were not relevant in supervised AV crash analysis. The first is crashes that occurred under driverless operation. Driverless crashes were studied in a different step, so we leave it out for this tabulation. We removed these events by manually reviewing the Accident Detailed Description field. The second category is crashes under conventional (manual) driving mode. Since we are using the Autonomous Mileage Reports from the DMV, we removed the crashes where the vehicle was manually driven. However, we identify and keep the occurrences where the autonomous system was disengaged right before the crash. In this case, the vehicle was operating autonomously until right before the crash and the autonomous system is responsible for the crash, so those cases need to be counted as automated collisions.

We also filtered for companies that are known to operate in San Francisco, namely Waymo, Cruise, Lyft (inactive since June 2021), and Zoox. Something to note here is that two of the data fields, whether the crash caused injuries and whether the crash involved vulnerable road users (VRUs), were incomplete from the checkboxes alone. We had to further consult the Accident Detail Description field to note the additional injuries or VRU involvement. Hence the final tallied results are higher than they would have been from counting the checkboxes alone.

**The CPUC TNC Annual Report, Fiscal Year 2020**

The CPUC TNC Annual Reports provide the crash and mileage data from human drivers for this comparison. The reports were submitted by Uber and Lyft. These reports were made available to the public after 2020 under Commission Decisions D.13-09-045 (20) and D.16-04-041 (21). According to an analysis by the San Francisco County Transportation Authority (SFCTA) (22), for the 2020 report, Uber had 99.99% complete required data, whereas Lyft was only 30% complete. In the most recent 2021 TNC report, however, both Uber and Lyft heavily redacted their Trips Requests data under Confidentiality Arguments. SFCTA found that both Uber and Lyft had 30% data completion (22). Because of the incomplete data for Lyft in 2020, and both Uber and Lyft in 2021, this analysis was constrained to focus on Uber's miles driven and crash data from 2020.

The CPUC Trips Requested Report in 2020 contains miles traveled and pickup and drop-off zip code for every trip that has been recorded on Uber. The total mileage traveled is calculated by summing the lengths of all trips that had starting and ending zip codes in the county of SF. The calculation is done with the help and data from SFCTA (22).

The CPUC Accident and Incident Report in 2020 contains the fields of data that are not redacted for each crash case, as shown in the left column of Table 1. From these data, we filter for crash cases in San Francisco by Incident Zip Code.

With both the mileage and crash data within the same time frame and same areas of operation, we can determine the rate of crashes per million miles for Uber human driving.

Table 1 – Data fields Used from CPUC TNC Accident and Incident Report and CPUC Quarterly Report on Driverless AV Operation

| CPUC TNC Accident and Incident Report | CPUC Quarterly Report on Driverless AV |
|---|---|
| • Submission Date<br>• Complaint ID<br>• Vehicle Make<br>• Vehicle Model<br>• Vehicle Year<br>• Incident Date<br>• Incident Zip Code<br>• Complaint Filed Date | • Year<br>• Quarter<br>• Collisions All<br>• Collisions Occupant<br>• Collisions Pedestrian<br>• Collisions Bicycle Scooter<br>• Collisions Motorcycles<br>• Collisions Other Vehicle |
| • Complaint Resolve Date<br>• Incident Type<br>• Incident Party<br>• Incident Claim<br>• Incident Description<br>• Primary Collision Factor | • Collisions Property<br>• Collision Severity Property Damage<br>• Collision Severity Injury Possible<br>• Collision Severity Injury Minor<br>• Collision Severity Injury Severe<br>• Collision Severity Fatality |

**Driverless AV Pilot Collision and Mileage Report from CPUC, March 2022 - July 2023**
Driverless AV Pilot mileage statistics are provided by the CPUC quarterly (19). Currently, two companies, Waymo and Cruise, have this permit in San Francisco. We can obtain the driverless mileage for both companies. Crashes under this permit are not reported to the CPUC, hence we have to utilize the data from NHTSA to identify these crashes.

**Driverless AV Deployment Collision and Mileage Report from CPUC, March 2022 - July 2023**
Driverless AV Deployment mileage and crash statistics are also provided by the CPUC quarterly (19). During this period only Cruise had driverless AVs providing fared rides in San Francisco. We obtained the driverless mileage for both companies, as well as the crashes with data fields that are relevant to our analysis, shown in the right column of Table 1.

**NHTSA Standing General Order on ADS Crash Report Data**
We obtained the NHTSA automated driving system crash data of all ADS's since June 2021 and identified a subset of data that we are interested in. The attributes we filtered on were: crashes when the ADS vehicle is driverless and is in San Francisco. This would allow us to obtain the driverless crash count for Waymo pertaining to their Driverless Pilot permit, since all the driverless crashes should be under this permit until July 2023. This would also allow us to calculate the driverless crash count from Cruise, who had both driverless pilot and deployment permits. However, as we know the driverless deployment crash count from CPUC, we can subtract that number to obtain the crash count under the Driverless Pilot permit.

## Results and Discussion

After aggregating the supervised AV crash reports in SF between December 2019 and November 2022, we have summarized the data in Table 2 below. From the 305 cases of DMV-reported crashes between Dec 2019 and November 2022, we identified 145 cases where the crashes happened within San Francisco and were operated under autonomous mode or disengaged immediately before the crash. This is compared to the autonomous mileage reports from the DMV for consistency in estimating the crash rate. The reports are from Waymo, Cruise, Lyft (inactive since June 2021), and Zoox.

Table 2 consists of categorical data fields from the report that are relevant to safety and environmental variables. Note that the categories with an asterisk (*) denote manual labeling using the Accident Description.

Table 2 - Supervised AV-Involved Crashes from December 2019 - November 2022

| Vehicle1 Damage | Count | Percentatge |
|---|---|---|
| Minor | 114 | 78.6% |
| Moderate | 17 | 11.7% |
| None | 11 | 7.6% |
| Major | 3 | 2.1% |

| Injury Level* | Count | Percentatge |
|---|---|---|
| No Injuries | 122 | 84.1% |
| Injuries Reported | 23 | 15.9% |

| Involved VRU(Ped/Cyclist/Scooter)* | Count | Percentatge |
|---|---|---|
| Involved VRU | 11 | 7.6% |
| Did Not Involve VRU | 134 | 92.4% |

| Collision Type AV | Count | Percentage |
|---|---|---|
| Not Available | 61 | 42.1% |
| Rear End | 42 | 29.0% |
| Side Swipe | 16 | 11.0% |
| Other | 9 | 6.2% |
| Head-On | 9 | 6.2% |
| Broadside | 7 | 4.8% |
| Hit Object | 1 | 0.7% |

| Collision Type Other Vehicle | Count | Percentage |
|---|---|---|
| Rear End | 57 | 39.3% |
| Head-On | 31 | 21.4% |
| Not Available | 28 | 19.3% |
| Side Swipe | 19 | 13.1% |
| Broadside | 7 | 4.8% |
| Other | 3 | 2.1% |

| Movement Vehicle 1 | Count | Percentatge |
|---|---|---|
| Stopped | 72 | 49.7% |
| Proceeding Straight | 33 | 22.8% |
| Slowing/Stopping | 14 | 9.7% |
| Making Left Turn | 7 | 4.8% |
| Making Right Turn | 6 | 4.1% |
| Backing | 4 | 2.8% |
| Changing Lanes | 3 | 2.1% |
| Stopped and Making Left Turn | 2 | 1.4% |
| Passing Other Vehicle | 1 | 0.7% |
| Other | 1 | 0.7% |
| Parking Manveuver | 1 | 0.7% |
| Not Available | 1 | 0.7% |

| Movement Vehicle 2 | Count | Percentatge |
|---|---|---|
| Proceeding Straight | 73 | 50.3% |
| Not Available | 10 | 6.9% |
| Changing Lanes | 12 | 8.3% |
| Making Left Turn | 7 | 4.8% |
| Slowing/Stopping | 6 | 4.1% |
| Stopped | 5 | 3.4% |
| Backing | 5 | 3.4% |
| Entering Traffic | 4 | 2.8% |
| Parked | 4 | 2.8% |
| Making Right Turn | 3 | 2.1% |
| Xing to Opposing Lane | 4 | 2.8% |
| Other | 2 | 1.4% |
| Other Unsafe Turning | 2 | 1.4% |
| Passing Other Vehicle | 3 | 2.1% |
| Travling Wrong Way | 1 | 0.7% |
| Making Left Turn + Unsafe Turning | 1 | 0.7% |
| Passing Other Vehicle + Travling Wrong Way | 1 | 0.7% |
| Right Turn + Passing Other Vehicle | 1 | 0.7% |
| Left Turn + Passing Other Vehicle | 1 | 0.7% |

| Autonomous Mode | Count | Percentatge |
|---|---|---|
| Autonomous | 119 | 82.1% |
| Disengaged and Manually Operated | 26 | 17.9% |

| Weather | Count | Percentage |
|---|---|---|
| Clear | 133 | 91.7% |
| Cloudy | 7 | 4.8% |
| Raining | 4 | 2.8% |
| Fog/Visibility | 1 | 0.7% |

| Lighting | Count | Percentage |
|---|---|---|
| Daylight | 99 | 68.3% |
| Dark with Street Lights | 40 | 27.6% |
| Dusk-Dawn | 5 | 3.4% |
| Dark with No Street Lights | 1 | 0.7% |

| RoadSurface | Count | Percentage |
|---|---|---|
| Dry and No Unusual Conditions | 135 | 93.1% |
| Wet and No Unusual Conditions | 4 | 2.8% |
| Not Available | 3 | 2.1% |
| Dry and Reduced Roadway Width | 2 | 1.4% |
| Dry and Obstruction on Roadway | 1 | 0.7% |

In terms of damage to the AV, the majority of the damages were minor damage (78.8%), with 11.7% being moderate damage, 7.6% no damage, and 1.7% major damages. The DMV crash reports frequently only describe damage to the AV, so the damages to the other vehicle or party involved remain unknown.

About 16% of AV crashes resulted in reported injuries. Three considerations apply here: first, the vehicles were mostly driven on local roads, so the speed is low. Lower speed would have a smaller likelihood of injury (23). The second is that there's a safety driver onboard, and they would be attentive to the vehicle's maneuvers, so the combination of a safety driver and the technology would allow the vehicle to minimize impact and avoid injuries; and third and most importantly, injuries may be underreported. Counts of injuries are inferred from each report's narrative section, and often the narrative only mentions injuries of people in the ADS-enabled vehicles, or the narrative does not mention injury. In these cases, it's not clear if others were injured.

The most common type of collision for the AV was rear-ended (29%), with sideswipe (11%) being the second most common. There are also 42% where the collision type is not reported. The most common types of collision for the other vehicle involved are also rear-end (39.3%) and head-on (21.4%).

In terms of autonomous modes, 82% of the crashes were under the autonomous mode, and 18% were under the conventional mode, but the autonomous system was disengaged right before the crash. These 18% could be categorized as autonomous mode because the underlying cause is the AV technology. The disengagement-before-crash cases were identified from the narrative description of the crash reports. We also would not know if there were cases when the safety driver disengaged the system and prevented the crash from happening, in which case there would not be a crash report.

When the crashes happened, half of the time, the vehicle was stopped. The next most frequent movements were proceeding straight (22.8%) and slowing down (9.7%). This does not reveal much about the safety of AV maneuvers. The next most frequent conditions were that it was either turning left (6.4%) or turning right (6.2%). However, we do not know how often the AVs make left and right turns, so we cannot comment on the relative safety of the two types of turns. There were also infrequent occurrences when the vehicle was backing up (3%), changing lanes (3%), or parking (<1%).

Some of the external factors associated with the crashes are – mostly in clear California weather (91.7% clear, with the other 4.8% cloudy, 2.8% rain, and <1% fog), during daylight (78.3%, with the rest being 21% night with streetlight, 3% dusk or dawn), and with a dry road (93.1%, with the rest being 3% wet roads and 4% obstructed roadway conditions). However, we don't have data to show the distribution of the weather and road surface conditions in which the companies were operating their vehicles to understand the relative exposure rates.

Finally, 11 crash cases involved vulnerable road users (pedestrians, bicyclists, or scooters, but not motorcycles) (7.6%). Note that this category does not include cases in which the AV crash may have been caused by a VRU, but the AV did not have contact with a VRU. Such a case would be an AV yielding to a pedestrian making an unexpected move by braking hard and causing another car behind to rear-end the AV.

We plotted the supervised AV miles driven by all AV testing companies in San Francisco between December 2019 and November 2022 with their monthly crashes in Figure 4.

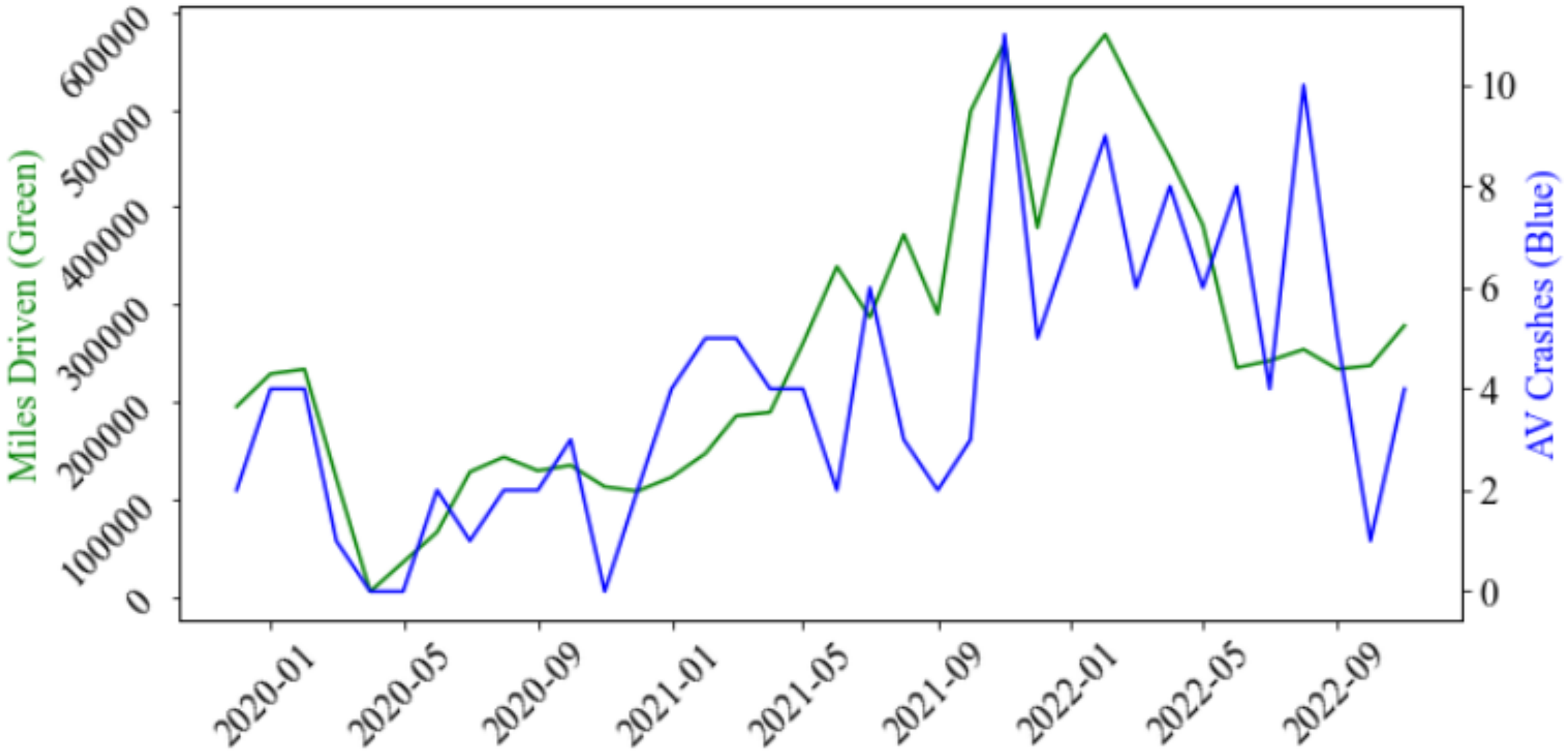

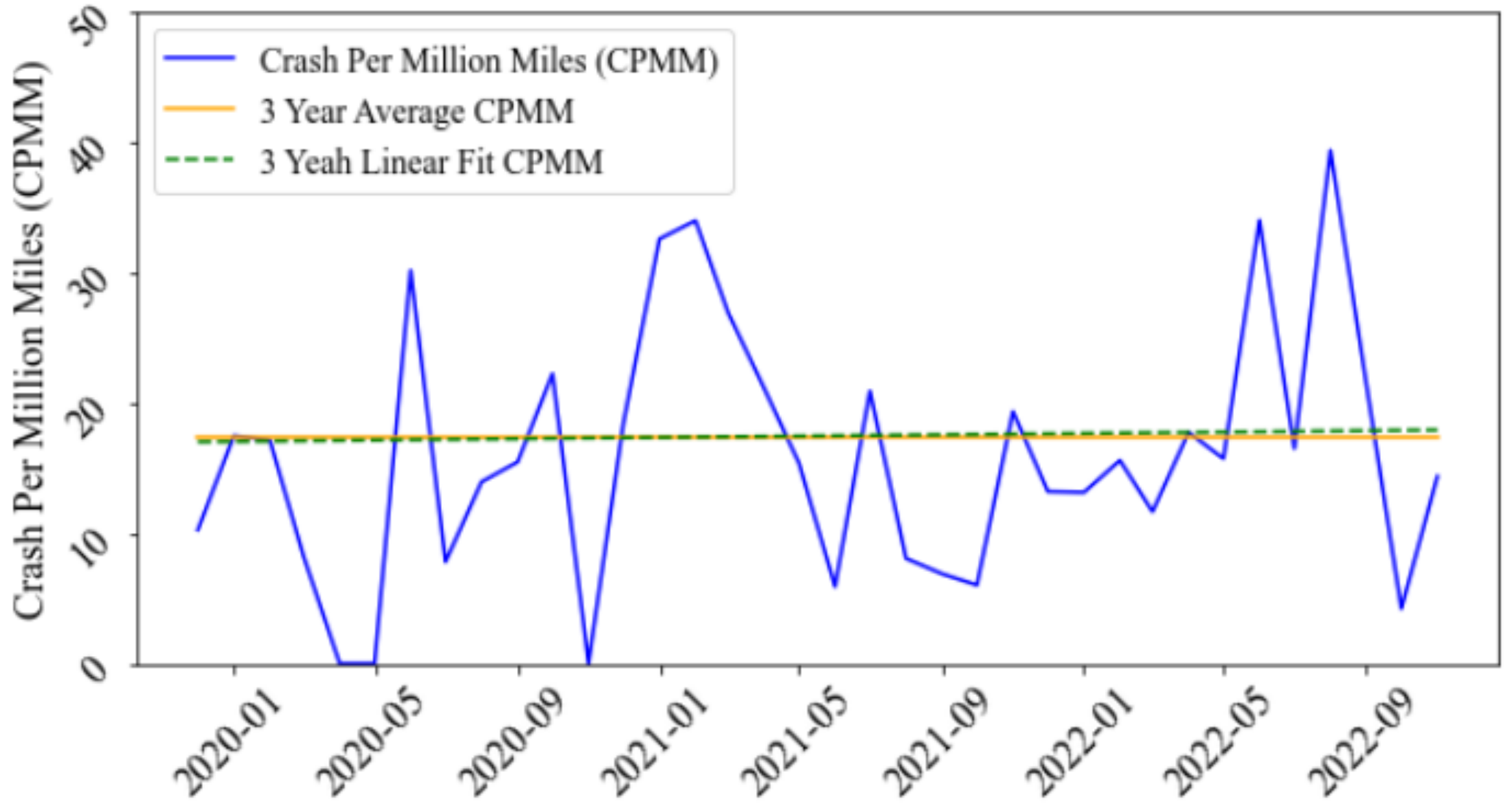

Figure 4 - Miles Driven Overlay with Crashes Over Time for Supervised AV from January 2020 to November 2022.

From the upper plot in Figure 4, we see a strong correlation between supervised AV miles driven and the number of crashes. Upon further analysis as shown in the lower plot, the crash per mile statistic fluctuates over time, with noticeable gaps in the graph during some months of 2020 when the amount of AV testing was severely diminished by COVID pandemic restrictions. Fitting a linear regression on the CPMM data over time, we see that the trend is in fact upwards, but only very slightly. It should also be recognized that this crash-per-mile statistic is an aggregate across all companies. It does not factor in the distribution of mileage by individual companies that were testing diverse systems of different levels of maturity at different times.

The aggregated categorical data in the Accidents and Incident Report (for human-driven operations) from Uber to the CPUC in 2020 are shown in Table 3 below.

Table 3 - TNC Reported Crash for human-driven Uber operations in San Francisco in 2020

| **Incident Type** | **Count** | **Percentage** |
|---|---|---|
| Multiple Vehicle Collision | 2015 | 93.4% |
| Open Door Into Vehicle | 68 | 3.2% |
| Pedestrian | 37 | 1.7% |
| Single Vehicle Collision | 28 | 1.3% |
| Struck Debris | 9 | 0.4% |
| Struck Animal | 1 | 0.0% |
| | | |
| **Incident Description** | **Count** | **Percentage** |
| Property damage alleged | 1635 | 75.8% |
| Property damage and bodily injury alleged | 502 | 23.3% |
| Bodily injury alleged | 16 | 0.7% |
| No property damage or bodily injury allged | 5 | 0.2% |
| | | |
| **Primary Collision Factor** | **Count** | **Percentage** |
| Claimant Primarily | 1183 | 54.8% |
| Driver Primarily | 494 | 22.9% |
| Undetermined | 481 | 22.3% |
| | | |
| **Characteristics from Incident Claims*** | **Count** | **Percentage** |
| Rear-end | 583 | 27.0% |
| Lane change/Intersection/Struck another car | 563 | 26.1% |
| Sideswipe | 332 | 15.4% |
| Unspecified | 259 | 12.0% |
| Backed-into | 146 | 6.8% |
| VRU | 135 | 6.3% |
| Open Door Into | 95 | 4.4% |
| T-Boned | 38 | 1.8% |
| Head-on | 7 | 0.3% |

The Characteristics from the Incident Claims category was derived from analyzing the narrative descriptions of each crash. In most cases, the crash narrative only discusses the main conflict, hence we identified likely mutually exclusive characteristics, such as collision with VRU and sideswipes and backed-into scenarios. We were able to categorize 88% of the crashes into pre-determined categories, as shown in Table 3 above.

Something to note here is that, although there is a category for colliding with pedestrians in the Crash Type, collisions with scooters and bicyclists fall under “Multiple Vehicle Collision”. Therefore, we had to identify collisions with VRUs through the narrative.

Using the mileage data and crash data for the following driving scenarios: human-operated Uber driving (from CPUC) and ride-hail datasets from the Cruise reports (from Cruise’s website), supervised AV driving (from DMV), driverless pilot operation by Waymo, driverless pilot by Cruise, and finally driverless deployment by Cruise (all from CPUC and NHTSA), we can compare the rate of crashes per million miles (CPMM). Figure 4 shows the result of the comparison.

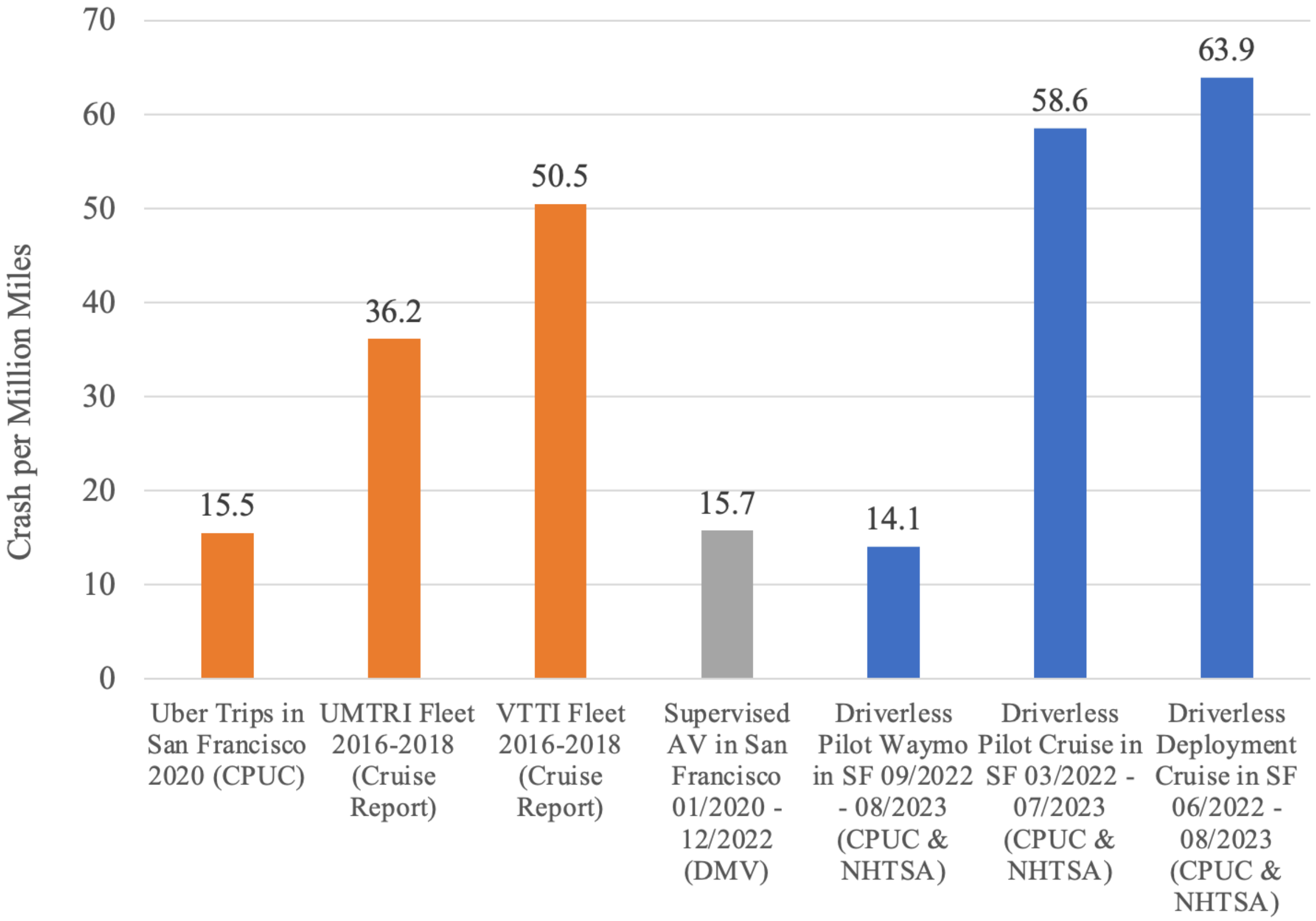

Figure 5 – Crash per Million Miles (CPMM) in San Francisco for Human-Operated Uber, Human-Operated Ride-hail Driving, Supervised AV Under Test, and Driverless AV Pilot and Deployments

Figure 5 shows the human-operated CPMM in orange, supervised AV CPMM in grey, and driverless AV CPMM in blue. We observe that the supervised AV's CPMM is almost equivalent to the crash rate for Uber drivers and is lower than both UMTRI and VTTI estimates of crash rates from their report to Cruise. We hypothesize that the supervised AV testing would include a mix of less mature systems as well as mature systems. This comparison is significant because it shows that the AV crash rate with a human safety driver, is similar to that of human drivers.

The three blue bars to the right in Figure 5 are the driverless CPMM under pilot or deployment permits for the initial months of operations from Waymo and Cruise as reported by them to CPUC. Waymo has reported to CPUC about one million driverless pilot miles, whereas Cruise has reported half a million driverless pilot and about a quarter million driverless deployment miles (details in Table 4 below). We separately calculated their individual CPMM. Waymo has a lower rate of CPMM compared to the human driving benchmark, as well as the aggregated supervised AV, indicating that their system may be safer under this metric. Cruise has a significantly higher CPMM, under both permits, when compared to the human benchmark and supervised AV driving. Its CPMM is also higher than Waymo's driverless CPMM.

We further compare crash characteristics among the TNC dataset, supervised AV, and driverless AV in Table 4 below. This table shows the differences in the order of magnitude of accumulated mileage for each category. The human-operated TNC mileage is an order of magnitude larger than the supervised AV mileage and is about two orders of magnitude larger than the driverless AV mileage in the data that are currently available. This means that with such a limited sample of driverless data, the analysis must remain tentative.

Table 4 - Comparison of crash characteristics among human-operated TNC, human-supervised AV, and driverless AV in San Francisco.

| | Uber Trips in San Francisco 2020 (SFCTA and CPUC) | UMTRI Fleet 2016-2018 (Cruise Report) | VTTI Fleet 2016-2018 (Cruise Report) | Supervised AV in San Francisco 01/2020 - 12/2022 (DMV) | Driverless Pilot Waymo in SF 09/2022 - 08/2023 (CPUC & NHTSA) | Driverless Pilot Cruise in SF 03/2022 - 08/2023 (CPUC & NHTSA) | Driverless Deployment Cruise in SF 06/2022 - 08/2023 (CPUC & NHTSA) |
|---|---|---|---|---|---|---|---|
| **Miles Driven** | 139,048,544 | 5,424,077 | 415,901 | 9,206,603 | 994,842 | 461,045 | 266,021 |
| **Crashes** | 2158 | 196.19 | 21 | 145 | 14 | 27 | 17 |
| **Crash Per Million Miles** | 15.5 | 36.2 | 50.5 | 15.7 | 14.1 | 58.6 | 63.9 |
| **Crash with Injuries** | 24.0% | N/A | N/A | 15.9% | 0.0% | 0.0% | 11.8% |
| **Crash Involving VRUs** | 6.3% | N/A | N/A | 7.6% | 0.0% | 0.0% | 5.9% |

From Table 4, we see that both supervised and driverless AVs have lower percentages of crash with injuries. The driverless pilot operation from both Waymo and Cruise did not report any crashes with injuries. However, the dataset for driverless operations is too small to make conclusive remarks, particularly because of the small sample of mileage for Cruise and the large difference between the CPMM indicated for Cruise and Waymo.

In terms of crashes involving vulnerable road users, the rates do not differ dramatically between the human, supervised AV, and driverless deployment operation of Cruise. The driverless pilot operations from both Waymo and Cruise also did not report any crashes with VRUs.

It is worth noting the significant difference in the incidence of injury crashes involving AVs compared with human-operated Uber trips. This is where we are more likely to see safety advantages from the AVs, since they are designed to be law-abiding and cautious drivers who do not violate traffic laws, and especially to avoid egregious violations that are more likely to cause injury crashes (speeding and red-light running). The data sample is still small, but it offers encouragement that as a larger body of data accumulates, we should start to see more convincing evidence of AV safety advantages.

## Conclusions and Future Study

This paper provides an initial comparison between supervised AV operation, driverless AV operation, and human-operated TNC data for similar operating conditions and geographical areas. We observe that the CPMM between supervised AV and human-operated TNC is similar. We also observe that Waymo has a lower CPMM with their driverless operation, whereas Cruise has a higher CPMM with their driverless operations.

By comparing the attributes of the AV crashes to the human-operated crashes, we see that both supervised and driverless AVs have fewer injury crashes, while the rates of crashes involving VRUs are similar. The sample size is still too small to draw definitive conclusions about AV safety, but the lower frequency of injury crashes offers encouragement about the potential for safety improvement.

The limitations of the analysis include:

1. The TNC companies (Uber and Lyft) and the automated vehicle developers (Waymo and Cruise) who provide data about their public operations to the CPUC have requested redaction of the large majority of the data that they deliver, and the CPUC has acquiesced to those redactions, severely limiting the range of data available for independent analysis. These redactions need to be lifted to enable more thorough analyses of the safety of both human and automated driving of ride-hailing services in San Francisco, to support better-informed public decision-making about these services.
2. The CPMM between CPUC's TNC report and the driverless AV crashes may not be completely comparable because NHTSA has a low severity threshold for crash reporting. All the AV crashes, regardless of damage level, would have to be reported to NHTSA. However, TNC drivers may not report all the crashes to their respective companies, though they are encouraged to. This means that the human driver benchmark may likely have a higher CPMM than shown in the data.
3. Uber's mileage and crash reports to the CPUC and the ride-hailing crash rate analyses supplied by Cruise as a human driving benchmark do not represent the safety of the overall population driving in San Francisco since the ride-hail drivers skew younger and more heavily male and drive longer hours, leaving them at higher risk.
4. The driving environments for the AV and human-operated TNC services in San Francisco are not exactly congruent, since Uber operates in SF all day, whereas driverless AVs, especially, have had limited day-time operation and more night-time operation. Running at night would mean a lower exposure of AVs to other road users and more difficult traffic conditions, giving them a traffic safety advantage.
5. There are inconsistencies in the Uber mileage reported to the CPUC in 2020, as indicated by the SFCTA report (20). This adds uncertainty to the estimate of their human-operated crash rate.

Future work should include integrating the current AV crash dataset with the NHTSA SGO records of ADS crashes, which contain more attributes, like roadway type and pre-crash speed, than the DMV reports. Follow-up studies should also be conducted with larger CPUC TNC datasets, which could become possible with a favorable decision on Rulemaking 12-12-011 (24), which would allow public sharing of TNC-reported data from 2014-2019.

## Acknowledgments

This research was sponsored by the State of California Transportation Agency, Department of Motor Vehicles (DMV). The contents of this paper reflect the views of the authors, who are responsible for the facts and accuracy of the data presented herein. The contents do not necessarily reflect the official views or policies of the State of California.

We'd like to thank the SFCTA for the analysis of Uber's vehicle miles traveled from the CPUC public 2020 TNC Annual Reports. The information was provided on 7/6/2023.